\documentclass[10pt,journal]{IEEEtran}
\usepackage[english]{babel}
\usepackage{amsmath,empheq}
\usepackage{cite}
\usepackage{algorithmicx}
\usepackage{algorithm}
\usepackage{algpseudocode}
\usepackage{tikz}
\usepackage{comment}
\usepackage{tikz-3dplot}
\usepackage{graphicx}
\usepackage[font=scriptsize]{subcaption}
\usepackage{breqn}
\usepackage{verbatim}

\usepackage{multirow}
\usepackage{xr}

\newcommand{\SSP}{\textit{shadow-inferred sun position}}
\newcommand{\MSP}{\textit{TL-inferred sun position}}
\newcommand{\framework}{\texttt{\textsc{ImageGuard}}}

\newcommand{\tool}{\textsc{\texttt{ImageGuard}}\xspace}

\algnewcommand\algorithmicinput{\textbf{Input:}}
\algnewcommand\INPUT{\item[\algorithmicinput]}

\algnewcommand\algorithmicoutput{\textbf{Output:}}
\algnewcommand\OUTPUT{\item[\algorithmicoutput]}

\usepackage{xspace}
\newcommand{\eg}{e.\,g.,\ }
\newcommand{\ie}{i.\,e.,\ }
\newcommand{\fig}{Figure\ }
\newcommand{\latinphrase}[1]{\textit{#1}}  
\newcommand{\etal}{\latinphrase{et~al.}\xspace}
\newcommand{\etals}{\latinphrase{et~al.}'s\xspace}

\makeatletter

\newcommand{\Rmnum}[1]{\expandafter\@slowromancap\romannumeral #1@}
\makeatother

\usepackage{url}
\makeatletter
\g@addto@macro{\UrlBreaks}{\UrlOrds}
\makeatother

\begin{document}

\title{Validating the Contextual Information of Outdoor Images for Photo Misuse Detection}


\author{\IEEEauthorblockN{Xiaopeng Li\IEEEauthorrefmark{1},
Xianshan Qu\IEEEauthorrefmark{1}, Wenyuan Xu\IEEEauthorrefmark{2}, Song Wang\IEEEauthorrefmark{1}, Yan Tong\IEEEauthorrefmark{1} and
Lannan Luo\IEEEauthorrefmark{1}}\\
\IEEEauthorblockA{\IEEEauthorrefmark{1}University of South Carolina, Columbia \quad
\IEEEauthorrefmark{2}Zhejiang University \\
}}


%
%

\maketitle


\begin{abstract}

The contextual information (i.e., the time and location) in which a photo is taken 
can be easily tampered with or falsely claimed by forgers to achieve malicious purposes, 
e.g., creating fear among the general public. A rich body of work has focused on detecting 
photo tampering and manipulation by verifying the integrity of image content. 
Instead, we aim to detect photo misuse by verifying the capture time and location of photos. 
This paper is motivated by the law of nature that sun position varies with the time and location, 
which can be used to determine whether the claimed contextual information corresponds with 
the sun position that the image content actually indicates. 
Prior approaches to inferring sun position from images 
mainly rely on vanishing points associated with at least two shadows, 
while we propose novel algorithms which utilize only one shadow in the image to infer the sun position. Meanwhile, we compute the sun position by applying astronomical algorithms which take as input the claimed capture time and location. Only when the two estimated sun positions are consistent can the claimed contextual 
information be genuine. 
We have developed a prototype called \tool. The experimental results show that our method can 
successfully estimate sun position and detect the time-location inconsistency with high 
accuracy. By setting the thresholds to be $9.4^\circ$ and $5^\circ$ for the sun position distance and the altitude angle distance, respectively, our system can correctly identify $91.5\%$ of falsified photos with fake contextual information.  

\end{abstract}




%
\IEEEpeerreviewmaketitle

\section{Introduction}
\label{sec:introduction}


With the rapid development of digital technologies and Internet, photos have become 
increasingly pervasive in our daily life for conveying information. For example, people use photos 
to express emotions, share ideas, and illustrate news stories on social media~\cite{imran2013extracting, Miller_2007, MALIK2016129}. 
However, not all of the photos are of good quality and used in proper ways. The distribution of fake photos on social media 
for malicious purposes is a growing concern~\cite{Boididou_2014}, which may undermine our trust in photography, interfere with law enforcement, and compromise national security, media, commerce, and more~\cite{seeing_is_believing, image_integrity}.   
Take the Hurricane Sandy happened at the northeastern U. S. in 2012 as an example. Numerous fake disaster photos and rumors were spread through social networks and caused panic and fear among the general public~\cite{gupta2013faking}. As a result, the U.S. Federal Emergency Management Agency has set up a ``rumor control'' section to defend against misleading information caused by fake photos on social networks~\cite{rumorControl}.

\begin{figure}[t]
  \graphicspath{ {./images/} }
  \centering
  \includegraphics[width=0.65\linewidth]{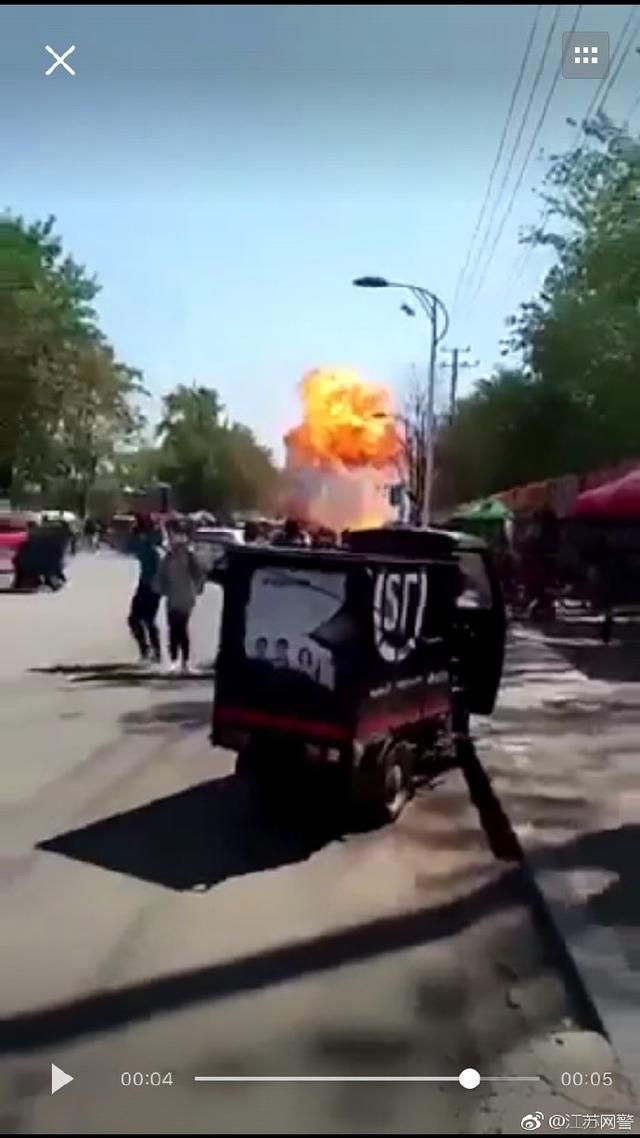} 
  \caption{The photo was captured in Baoding China in April 2017, but it was misused for a news event happened in Xuzhou China on June 15, 2017.}
  \vspace{-2pt}
  \label{fig:fake_explosion}
\end{figure}

Modifying the content of an image is one of the most common methods to create 
fake photos; thus, most existing techniques focus on identifying falsified content 
within a photo to detect whether or not the photo is fake (or tampered)~\cite{fridrich-soukai-lukas-dfrw-03, Bayram09anefficient, conf/icassp/PanL10, 1381775, mediaforensics/CarvalhoFK15, Kee:2014}. 
For example, some approaches are proposed to detect copy-move manipulation in digital 
images~\cite{fridrich-soukai-lukas-dfrw-03, Bayram09anefficient, conf/icassp/PanL10}, 
and some others are proposed to detect photo tampering by leveraging shadows and lighting 
of images~\cite{mediaforensics/CarvalhoFK15, Kee:2014}.

However, sometimes an adversary does not modify the content of a photo, but instead 
\emph{directly misuses the photo and claims the photo is taken at a different time 
or location}, to achieve malicious purposes, such as spreading misleading information, 
attacking and deceiving others, or generating chaos and panic. In such a case, existing
countermeasures would fail to detect such fake photos as their content is intact. 
Consider the example of Figure \ref{fig:fake_explosion} which shows a 
clip of an explosion video\footnote{http://www.twoeggz.com/news/1949256.html}. 
This video was spread on social networks and claimed to be associated to the explosion occurred in Xuzhou, China on June 15th, 2017. However, it was actually 
captured in Baoding, China in April 2017. This example indicates a critical need 
for computer-aided, automated photo verification techniques that are capable of validating 
the capture locations (i.e., GPS coordinates) and times of photos to prevent photo misuse.


Validating the capture time and the location of a photo is a promising yet challenging task. 
Although most photos taken by modern cameras have timestamps and GPS information embedded 
in its metadata, simply relying on the metadata to determine when and where the photo has 
been taken is unreliable as the metadata can be easily altered by powerful digital media 
editing tools~\cite{kee11, KakarS12a}. Another possible way is to recognize the landmarks in photos and 
use the landmarks to validate the place where the photo has been taken. However, most image 
scenes, such as public lawns, parking lots, beaches and roadsides, tend to be similar in 
many different places and, hence, this is also infeasible. 
Furthermore, objects such as clothing and colors of trees can imply the time when the photo has 
been captured, but they can only reveal a relatively long time span (\eg a T-shirt is suitable 
from April through October in many places), and thus cannot be used to infer a precise or 
relatively short capture time of the photo.


Our observation is that shadows cast by the sun vary over time and location; that is, the shadow's orientation and the length ratio between the vertical object and its shadow are directly affected by the position of the sun (\ie the orientation and height), which changes with time. Moreover, even at the same time, the shadow's orientation and the length ratio can still be various in two different cities, as the sun position also depends on the GPS coordinates. Therefore, \emph{the shadows of vertical objects are determined by the sun position, which is affected by the time and location.} 

Inspired by the observation, we propose to use shadows that appear often in outdoor images to verify the capture times and locations of images. Specifically, we first use the shadows of vertical objects in an image to estimate the sun position. We then compute another estimation of the sun position using the capture time and location 
of the image, which are obtained from the image metadata or claimed by the photo's user. Note that we assume an adversary, if wants to misuse a photo, tends to modify the time and location in the metadata to align the metadata with those she claims. Finally, if the difference between the two estimations of the sun position is above the set threshold, we determine that the photo is misused.

While it is easy to measure the length ratio of a vertical object and its shadow in real world
(then the length ratio can be used to compute the sun position), it is difficult to obtain such 
information from a 2D image. The reason is that the projection from a 3D scene to a 2D image is a 
non-linear transformation; as a result, the length ratio measured on the image could be significantly 
different from the true value in real world.
In addition, the length ratio measured on the image depends not only on the scene in real world 
but also on the camera's intrinsic parameters and viewing angle. Determining the shadow's orientation 
from a 2D image is also challenging due to a lack of the third dimension information. 
Although single view reconstruction has been 
extensively studied in the field of computer vision, there is no generalized way to recover the 
relative positions of objects from a single image. Lalonde \etal analyze the distributions 
of natural illumination from images, and estimate the sun position based on a proposed probability 
model~\cite{Lalonde2009, Lalonde2010}; however, the results 
show that the error is larger than $22.5^{\circ}$ for half of the testing photos when predicting 
the sun's position. Junejo \etal propose an approach that uses two vertical objects' 
shadows in the image to estimate the sun position~\cite{geotempestimation}; however, sometimes 
only one vertical object may exist in images.

By exploring the camera model and the geometry relations among the camera, the objects and their shadows, 
we devise new algorithms to estimate the sun position based on only one vertical object and its shadow. 
Specifically, we first investigate how a camera object pointing in 3D world is projected onto a 2D image, 
and then analyze the geometry constraints between the interested object, its shadow and the ground. Based 
on the relations, we recover the 3D positions of the object and its shadow, which can be used to infer the 
sun position.

We have developed a prototype, called \framework{} to verify the Time-Location 
of outdoor images. We collected around 200 photos across three different countries and 16 cities. 
We conducted three sets of experiments and analyzed the validation performance under different variables 
and thresholds. \framework{} achieves a best performance on detecting the falsifications of the time of a day. Overall, using the distance thresholds of sun position and sun altitude angle, we can correctly recognize $92.5\%$ of all the true samples with an error of $8.5\%$ in identifying false samples. Our experimental results show that \framework{} 
can successfully estimate sun position and verify the time-location of outdoor images with high accuracy. 

\vspace{2pt}
\textit{\textbf{Extension.}} This paper extends our previous work~\cite{previous_work}, in the following perspectives.

\begin{itemize}
\item New algorithms (Section \ref{sec:estimate_sun_position}) are proposed which need only one shadow and the camera can be tilted downwards or upwards, while the previous algorithms need either two shadows or one shadow with the camera held vertically.
\item To examine the new algorithms, the previous dataset is enlarged by adding new images that were captured with cameras tilted at different angles (Section \ref{sec:dataset2}). Moreover, a synthetic dataset (Section \ref{sec:dataset1}) is added to examine the noise resilience of the new algorithms and analyze how the algorithms perform in different situations (Section \ref{sec:algo_evaluation}). 
\item A detailed discussion on the attacks against our framework is added (Section \ref{sec:attacks_on_framework}).
\end{itemize}

\vspace{2pt}
\textit{\textbf{Scope of Our Paper.}} 
We focus on the validation of the time-location consistency of outdoor images to
detect image misuse. That is, \tool\ will provide a Yes/No answer to the question: 
\emph{is the image misused?} or, \emph{is the time or location the image being taken 
consistent to that the image being used?} As \tool\ uses shadows in an image to verify 
the time and location, the outdoor image should contain at least one shadow. 
Moreover, we assume the adversary does not tamper with the content of the image 
(or at least the shadow part of the image); if the image has been modified, many 
existing techniques can be adopted to detect the fake image~\cite{fridrich-soukai-lukas-dfrw-03,Bayram09anefficient,conf/icassp/PanL10,mediaforensics/CarvalhoFK15,Kee:2014}, which is out of the scope of our work.

\vspace{2pt}
\textit{\textbf{Contributions.}} We make the following contributions. 
\begin{itemize}
\item 
We propose a novel approach that relies on the variances of the sun position 
for validating the time and location of outdoor images to detect image misuse. 
\item
We demonstrate that the sun position can be acquired from shadows and design algorithms to estimate the sun position from one vertical object and its shadow in the image, without the constraint of camera tilting angles. 
\item 
We implement the prototype and evaluate it using a dataset containing 200 
photos collected in 16 cities across China, the U.S. and Japan. The experimental results show that
\tool\ can successfully estimate the sun position and detect the time-location inconsistency
with high accuracy, and is effective and efficient.
\end{itemize}


\section{Background}
\label{sec:background}

\begin{figure}[t]
  \graphicspath{ {./images/} }
  \centering
  \includegraphics[width=0.76\linewidth]{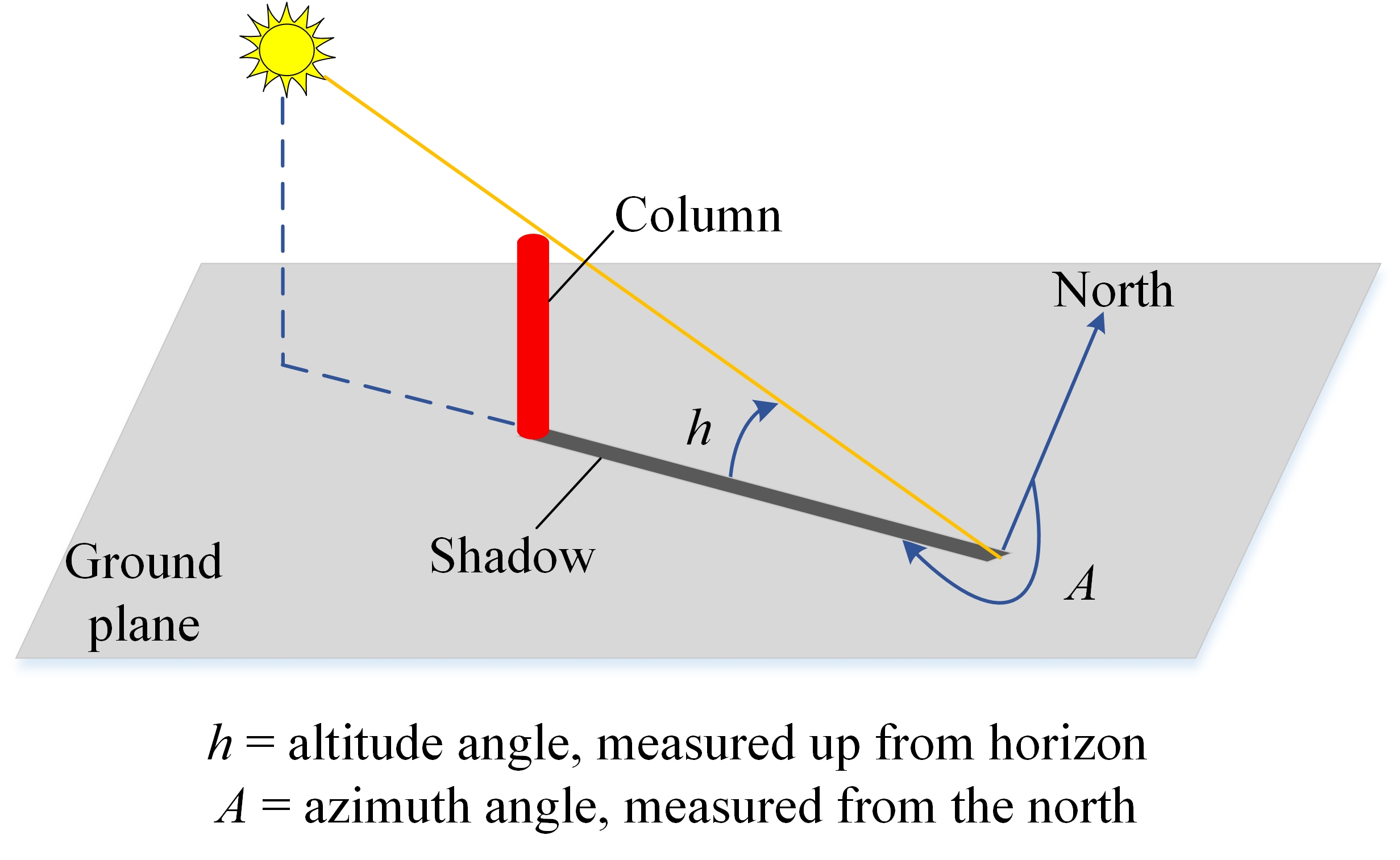}
  \caption{An illustration of the altitude and azimuth angles of the sun.}
  \label{fig:sun_position}
\end{figure}

\begin{figure*}[t]
\graphicspath{ {./images/} }
\centering
\begin{subfigure}[b]{0.30\textwidth}
  \includegraphics[width=\textwidth]{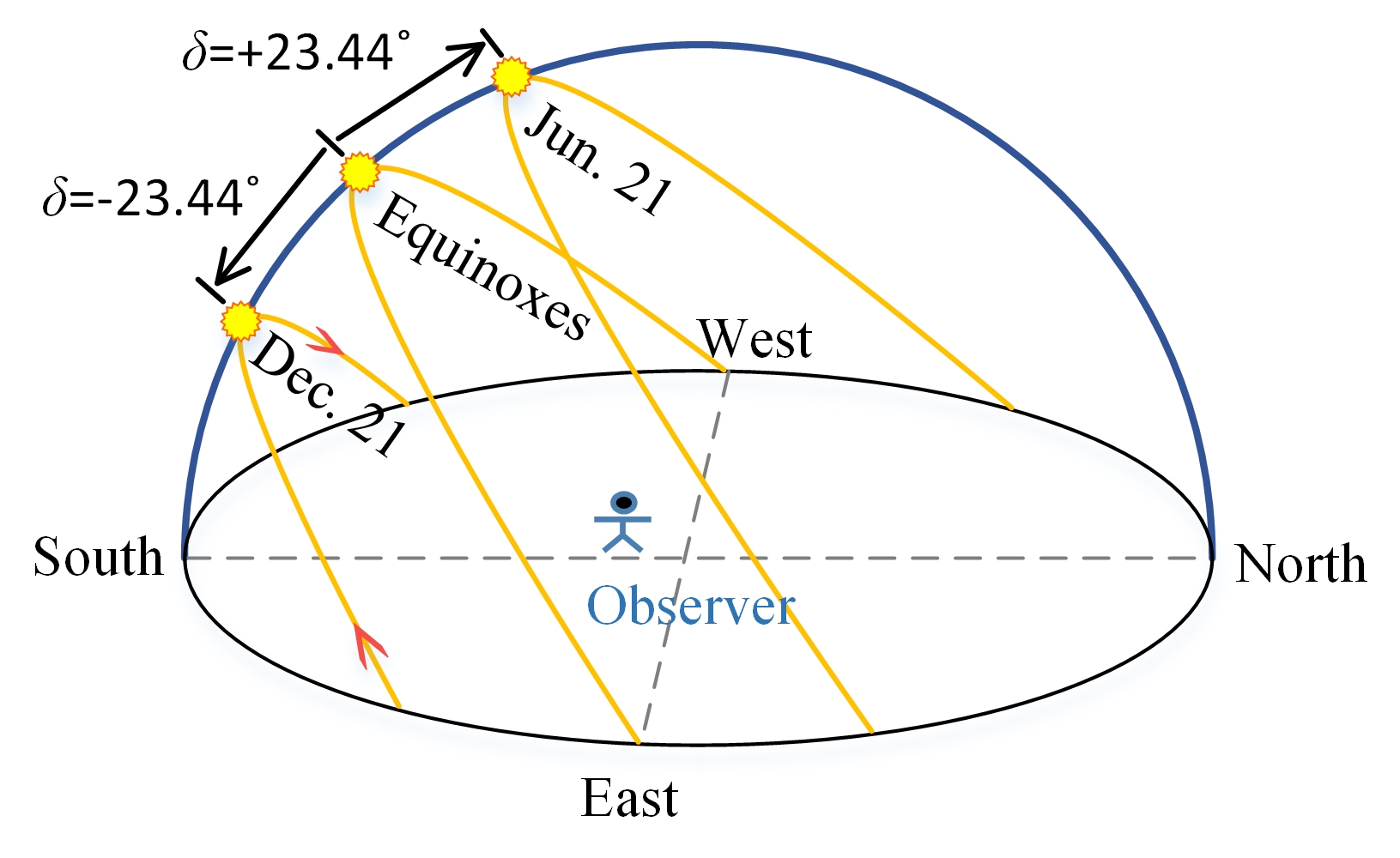}
  \caption{The path of the sun across the sky as observed on various dates in the northern hemisphere.}
  \label{fig:sun_path0}
\end{subfigure}
\hfill
\begin{subfigure}[b]{0.30\textwidth}
	\includegraphics[width=\textwidth]{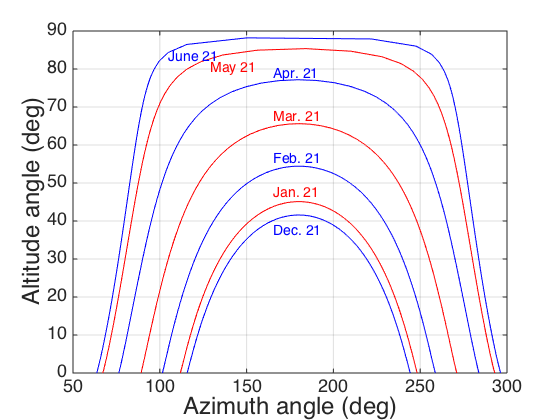}
	\caption{at $25^{\circ}$ north latitude in the U.S.}
	\label{fig:sun_path1}
\end{subfigure}
\hfill
\begin{subfigure}[b]{0.30\textwidth}
	\includegraphics[width=\textwidth]{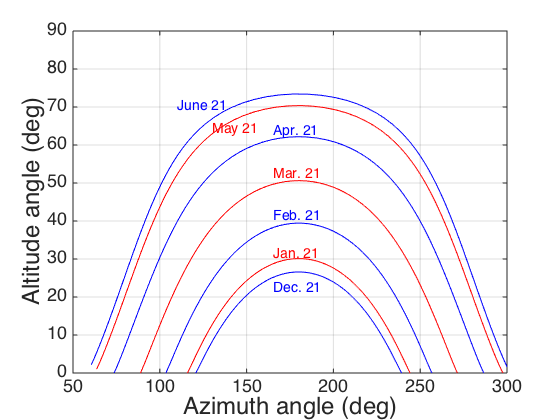}
	\caption{at $40^{\circ}$ north latitude in the U.S.}
	\label{fig:sun_path2}
\end{subfigure}

\caption{The same path of the sun observed at two latitudes.}
\label{fig:sun_path}
\end{figure*}

\subsection{Sun Position Definition}
The position of the sun in the sky is defined by an azimuth angle and an altitude angle. 
An azimuth angle describes the direction of the sun, and an altitude angle defines 
the height of the sun \cite{Sundials}. As shown in Figure~\ref{fig:sun_position}, 
the sun azimuth angle $A$ is measured clockwise in the horizontal plane, from the 
north to the direction of the sun. It varies from $0^{\circ}$ (north) through 
$90^{\circ}$ (east), $180^{\circ}$ (south), $270^{\circ}$ (west), and up to 
$360^{\circ}$ (north again). The altitude angle $h$ is measured from the horizontal 
to the sun and it thus ranges from $-90^{\circ}$ (at the nadir) through $0^{\circ}$ 
(on the horizon), up to $90^{\circ}$ (at the zenith). For instance, when the sun 
crosses the meridian, its azimuth is $180^{\circ}$ and altitude is at its largest 
value in a day.

\subsection{Motions of the Sun}
Observed from any location on the earth, the sun moves continuously across the sky throughout days and years. The position of the sun can be completely different observed at different time and places.

\textbf{Daily Path.} Because of the earth's daily rotation, the sun appears to move along with the celestial sphere every day. It makes a $360^{\circ}$ journey around the celestial sphere every 24 hours. Figure~\ref{fig:sun_path0} shows three of the sun's daily paths viewed on the earth. To an observer on the earth, the sun rises somewhere along the eastern horizon, and goes up to the highest point (zenith) around the noon, then goes down until it sets along the western horizon.  Accordingly, the cast shadows of any objects move oppositely from somewhere along the west to somewhere along the east. Their lengths vary with the sun's altitude angle. The higher the sun is above the horizon, the shorter the shadow is. Thus, the shadow that a camera takes at different times of the day will be different.
 
\textbf{Yearly Path.} The sun's daily path across the sky also changes throughout the year. The reason is that the earth does not rotate on a stationary axis and the tilt in the axis varies each day with respect to the earth's orbit plane. Observing on the earth, the sun looks higher in the summer than in the winter at the same time in the day. As shown in Figure~\ref{fig:sun_path0}, the sun follows different circles at different days in one year: most northerly on June 21st and most southerly on December 21st. The sun's motion along the north-south axis over a year is known as the declination of the sun, denoted by $\delta$. Thus, the sun position inferred from photos taken at the same location and time but different days in a year will be different due to the sun's declination.

\textbf{Different Latitudes.} As the sun travels across the sky, the observed altitude angle varies according to the latitude of the observer. The further north or south we go from the equator, the lower the sun's altitude becomes. Figure~\ref{fig:sun_path1} and Figure~\ref{fig:sun_path2} show the sun's altitude angle versus the azimuth angle observed at $25^{\circ}$ north latitude and $40^{\circ}$ north latitude respectively. The sun's altitude angle observed at $25^{\circ}$ north latitude is higher than the altitude angle observed at $40^{\circ}$ north latitude at the same time. Thus, the sun position inferred from photos taken at the same time but different latitudes will be different.
\section{Overview}
\label{sec:overview}
We specify the threat model, overview the framework of \framework{}, and describe the design assumptions in this section. 

\begin{figure*}[t]
  \graphicspath{ {./images/} }
  \centering
  \includegraphics[width=0.96\linewidth]{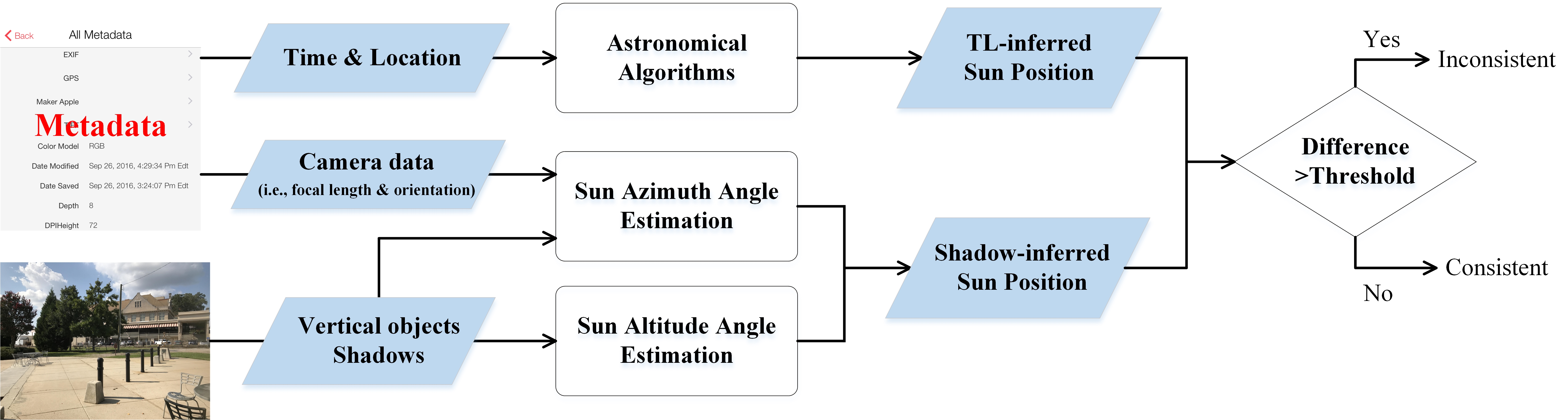}
  \caption{The work flow of the proposed \framework{} framework.}
  \label{fig:workflow}
\end{figure*}

\subsection{Threat Model}

We assume that an attacker modifies the capture time and location embedded in an image's metadata to achieve malicious goals, but does not tamper with or manipulate the content of the image, especially the objects and shadows in the image. Note that even if she modifies the image, we can still detect it utilizing the prior works \cite{fridrich-soukai-lukas-dfrw-03, Bayram09anefficient, conf/icassp/PanL10, 1381775, mediaforensics/CarvalhoFK15, Kee:2014}. The attacker's motivation could be creating an atmosphere of fear during an occurred social event, or providing false evidence to a court of law.


We mainly consider two possible attack methods. First, we assume that the image being exploited have complete metadata enclosed. 
In order to convince people that image content was captured at her claimed time and location, the attacker tends to modify the original metadata accordingly or generate new metadata. Modifying the capture time and location enclosed in metadata is quite easy to accomplish by using metadata editing tools~\cite{kee11, KakarS12a}, such as ExifTool~\cite{ExifTool}. 

Second, the attacker may remove the metadata from the image, then she can of course claim any time and location she wants. In such case, 
people cannot validate the attacker's claim immediately by checking the metadata. Utilizing our framework, we can still detect the inconsistency between her claim and the actual time and location.


\subsection{Overview of $\protect\framework$}

Our goal is to validate whether the claimed capture time and location of an image are true.

\textbf{Basic Idea.} Although an attacker can modify the metadata and claim that a photo was taken at time $X$ and location $Y$, she won't be able to change the ``time'' and ``location'' information that is embedded in the content of an image. Motivated by the observation that shadows are determined by the position of the sun varying with time and location, we use shadows that appear often in outdoor images to obtain the unchangeable information associated with time-location. Specifically, on one hand, we utilize the contents in images\textemdash{}vertical objects and resulting shadows\textemdash{}to extract the sun position. On the other hand, we utilize the claimed capture time and location to obtain a second estimation of the sun position. If these two estimations are close enough, we consider the capture time and location to be true with a high probability. 

\textbf{Workflow.} Figure \ref{fig:workflow} shows the workflow of our approach. For convenience of description, we use the term $\MSP$ to refer to the sun position calculated from claimed capture time and location, and use the term $\SSP$ to refer to the sun position estimated from shadows in the image. In this paper, the \emph{camera data} mainly represents the camera's focal length and orientation (i.e., the camera's pitch and yaw). Vertical objects refer to the ones that are perpendicular to the ground plane.

Our framework works as follows. To compute the $\MSP$, we collect the claimed capture time and GPS information from the image metadata and then apply astronomical algorithms. To calculate the $\SSP$, we extract the objects and the resulting shadows from the image using Matlab toolbox. We then run our algorithms based on camera projection model to recover the real positions of the objects and their shadows in real world, the sun position is computed according to the spatial geometrical relations between the objects and their shadows. In particular, the \emph{camera data} is needed for estimating the sun's azimuth angle, while it is not necessary for estimating the sun's altitude angle. 

Once the two estimations of the sun position are obtained, we determine the differences between these two estimations from three aspects and compare them with the selected thresholds. If the differences are smaller than the corresponding thresholds, we consider these two estimations are close enough indicating the same sun position, and hence the claimed capture time and location are true. Otherwise, they are considered to be falsified. Note that \textit{capture time} in this paper denotes the date of year and the time of day unless otherwise indicated.

\textbf{Assumption.} Without loss of generality, we assume that the camera has zero skew and square pixels (\ie unit aspect ratio), and it has zero roll angle, which corresponds to the typical situation where the camera looks at the front without tilting to left or right. Note that the camera is free to be tilted upwards or downwards. We further assume that the ground where the interested shadows rest on is approximated to level. Of course, at least one vertical object and its shadow have to be visible in the image. The objects can be human beings, road signs, lampposts, tree trunks and so on.

\section{Shadow-inferred Sun Position}
\label{sec:estimate_sun_position}
This section first provides an introduction to camera projection model, based on which we design the algorithms for estimating sun altitude and azimuth angles from one shadow in the image.

Throughout this section, we use the following notation conventions. A 3D point is denoted by a capital and bold letter ($\mathbf{X}$) and its projection on the 2D image is denoted by a lowercase and bold letter ($\mathbf{x}$). We also define $\mathbf{X} = [X_i, Y_i, Z_i, 1]^T$ and $\mathbf{x} = [x_i, y_i, 1]^T$ in homogeneous coordinates. A 3D vector is referenced by $\overrightarrow{\mathbf{X}_i\mathbf{X}_j}$. We refer to a vertical object in the world and its projection in the image as italicized letters $O_i$ and $o_i$ respectively. In addition, capital letters in roman font ($\mathrm{M}$) denote matrices.

\subsection{Camera Projection Model}
\label{sec:camera_model}
\begin{figure}[t]
  \graphicspath{ {./images/} }
  \includegraphics[width=0.85\linewidth]{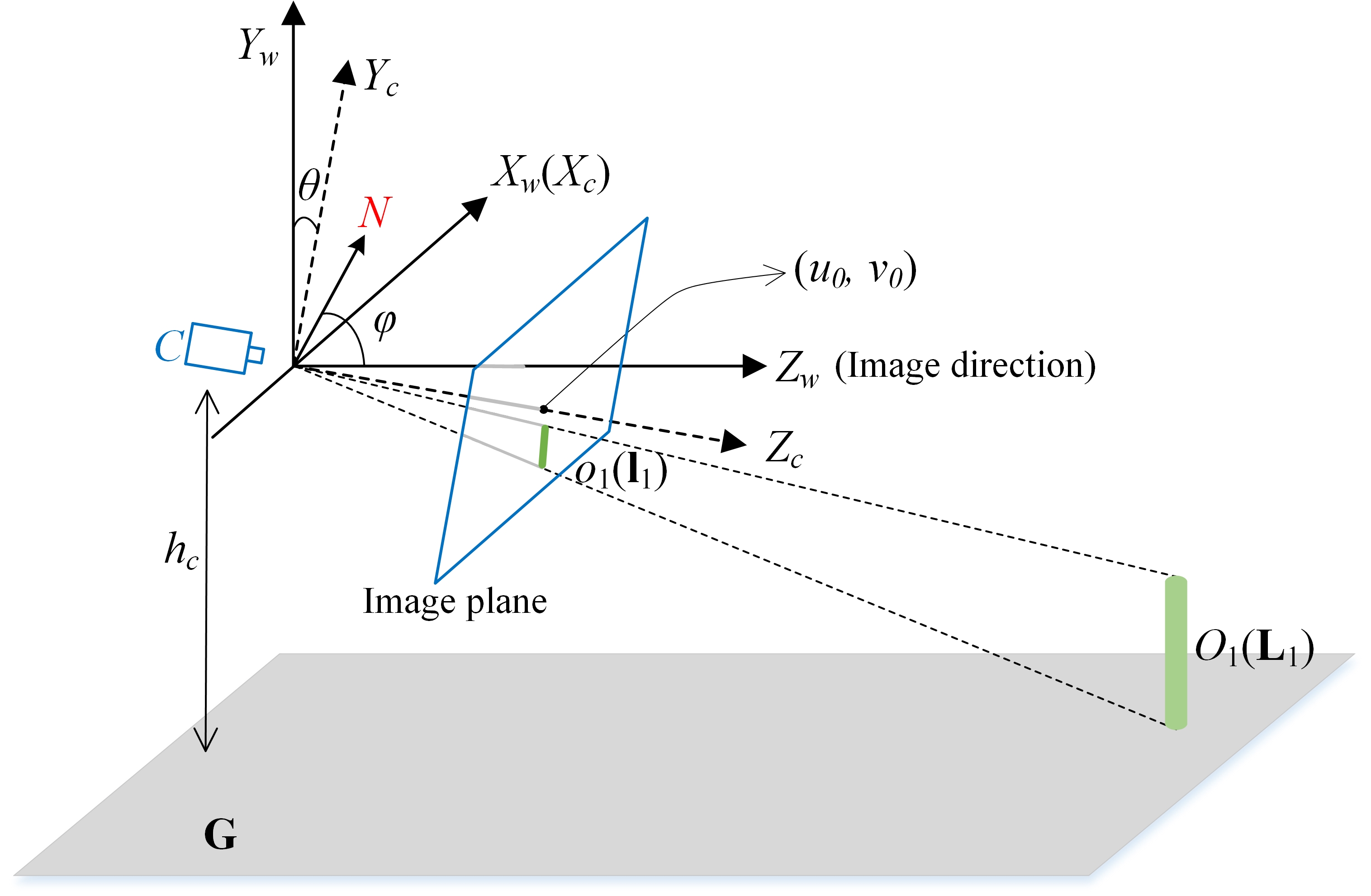}
  \centering
  \caption{The camera projection model. The camera reference frame is denoted by $(\mathrm{X}_c, \mathrm{Y}_c, \mathrm{Z}_c)$ and centered at the world reference frame $(\mathrm{X}_w, \mathrm{Y}_w, \mathrm{Z}_w)$, where $\mathrm{X}_w$ coincides with $\mathrm{X}_c$ and $\mathrm{Y}_w$ is vertical. The camera has inclination angle $\theta$ w.r.t. vertical and azimuth angle $\varphi$ w.r.t. North ($N$). We assume that it has no roll angle (i.e., $\mathrm{X}_c$ is parallel to the ground level). The vertical object $O_1$ in the 3D world is projected at $o_1$ in the image.}
  \label{fig:camera_model}
\end{figure}

Figure \ref{fig:camera_model} illustrates the projection model of a pinhole camera, where the camera $C$ has coordinate frame $(\mathrm{X}_c, \mathrm{Y}_c, \mathrm{Z}_c)$, and its center coincides with the origin of the world coordinate frame $(\mathrm{X}_w, \mathrm{Y}_w, \mathrm{Z}_w)$. We define $\mathrm{Y}_w$ to be the vertical direction and the $\mathrm{X}_w\mathrm{Z}_w$ plane to be parallel to the ground plane. We represent camera orientation $CO=\{\theta, \varphi\}$ using two parameters, where $\theta$ denotes the camera's inclination angle (\ie pitch) with respect to vertical and $\varphi$ denotes the camera's azimuth angle (\ie yaw) with respect to North. The angle $\varphi$ indicates the direction the camera is facing (i.e., the image direction). Both angles are included in the image's metadata.

To express the mapping from the 3D world to a 2D image, we model the projection of a pinhole camera as $\mathbf{x}=\mathrm{P}\mathbf{X}$ from
the point of view of projective geometry. $\mathrm{P}$ is called the homogeneous camera matrix and written as
\begin{equation}
\label{eq:camera_matrix}
	\mathrm{P} = \mathrm{KR}[\mathrm{I}|\mathbf{0}]\,,
\end{equation}
where $\mathrm{K}$ is the camera intrinsic matrix, $\mathrm{R}$ relates the camera rotation with respect to the world coordinate frame, $\mathrm{I}$ is a 3x3 unit matrix, and $\mathbf{0}$ is a zero vector and indicates that the camera is centered at the world coordinate frame.

The matrix $\mathrm{K}$ in Eq. \ref{eq:camera_matrix} can be given by
\begin{equation}
\label{eq:camera_internal_matrix}
	\mathrm{K} =
\begin{bmatrix}
f & 0 & u_0\\
0 & f & v_0\\
0 & 0 & 1
\end{bmatrix}\,,
\end{equation}
which assumes that the camera has zero skew, the intersection of the optical axis and the image plane is at the center of the image, and the pixels are square. Such assumptions are true for current camera technologies \cite{GeoCV, Wu2010}. In Eq. \ref{eq:camera_internal_matrix}, $(u_0,v_0)$ denotes the coordinates of the image center, and $f$ indicates the camera's focal length which is either included in the image metadata or inferred using algorithms in~\cite{zhenyou_cal0, zhenyou_cal1, geotempestimation}  

To determine the rotation matrix $\mathrm{R}$, let's say the camera rotates by $\theta$ (i.e., the inclination angle) along the axis $\mathrm{X}_w$ w.r.t the world frame. Then we can write $\mathrm{R}$ as follows:
\begin{equation}
\label{eq:rotation_matrix}
	\mathrm{R} =
\begin{bmatrix}
1 & 0 & 0\\
0 & \cos\theta & -\sin\theta\\
0 & \sin\theta & \cos\theta
\end{bmatrix}\,,
\end{equation}
where the angle $\theta$ can be computed by camera inertial sensors. Without loss of generality, we assume that a photographer normally holds the camera with an inclination in the range of $(-45^{\circ}, 45^{\circ})$. This, of course, is consistent with our common habits of photographing. Finally, we have the mapping matrix $\mathrm{P}$ as follows:
\begin{equation}
	\mathrm{P} =
\begin{bmatrix}
f & u_0\sin\theta & u_0\cos\theta & 0 \\
0 & f\cos\theta+v_0\sin\theta & v_0\cos\theta-f\sin\theta & 0 \\
0 & \sin\theta & \cos\theta & 0
\end{bmatrix}\,.
\end{equation}

\begin{figure}[t]
\graphicspath{ {./images/} }
\centering
\begin{minipage}[t]{.45\textwidth}
  \centering
  \includegraphics[height=.60\linewidth]{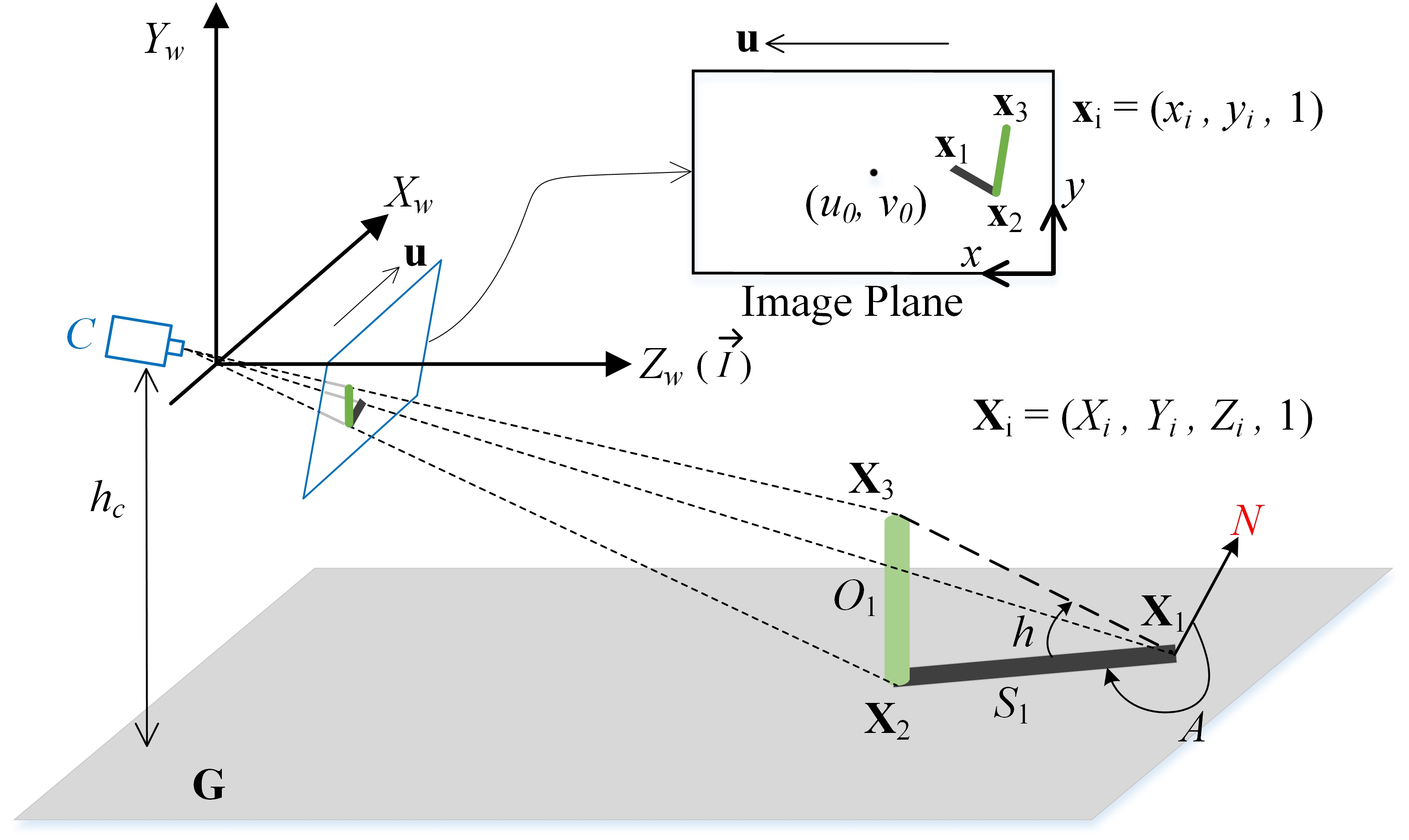}
  \captionof{figure}{Estimate the sun's altitude and azimuth angle with one shadow.}
  \label{fig:one_shadow_scenario}
\end{minipage}
\end{figure}

\subsection{Estimating Sun Altitude Angle}
\label{sec:altitude_measurement}

To estimate the sun altitude angle, we consider the scenario where only one vertical object and its shadow are visible in the image. Figure \ref{fig:one_shadow_scenario} depicts how the object $O_1$ and its shadow $S_1$ are projected onto a 2D image by the camera $C$ with an inclination angle $\theta$. 
The intuition of this algorithm is to recover the positions (\ie coordinates) of $\mathbf{X}_1$, $\mathbf{X}_2$ and $\mathbf{X}_3$ based on the camera model and the resulting projections. 

To recover the coordinates of $\mathbf{X}_1$ and $\mathbf{X}_2$, we explored two constraints. First, considering the fact that $\mathbf{X}_1$ and $\mathbf{X}_2$ lie on the ground plane, they satisfy $\mathbf{X}_i^T \mathbf{G} = 0$ for $i = 1, 2$ and $\mathbf{G}=[0, 1, 0, h_c]^T$. Second, these two points in space are mapped to the image according to the camera's mapping matrix, so $\mathbf{x}_i = \mathrm{P}\mathbf{X}_i$. By solving the above equations, we can obtain the coordinates of points $\mathbf{X}_1$ and $\mathbf{X}_2$. Then we have the vector $\overrightarrow{\mathbf{X}_1\mathbf{X}_2}$.

To determine the coordinates of $\mathbf{X}_3$, we also utilize two constraints. First, since the object $O_1$ is perpendicular to the ground plane ($\mathrm{X}_w\mathrm{Z}_w$), the point $\mathbf{X}_3$ has the same $X$ and $Z$ components with $\mathbf{X}_2$. Second, the projection of $\mathbf{X}_3$ from space onto the image also satisfies $\mathbf{x}_3 = \mathrm{P} \mathbf{X}_3$. Using these two constraints, we can obtain the coordinates of $\mathbf{X}_3$. Once the coordinates of $\mathbf{X}_1$ and $\mathbf{X}_3$ are obtained, we can calculate the vector $\overrightarrow{\mathbf{X}_1\mathbf{X}_3}$.

Finally, the angle between $\overrightarrow{\mathbf{X}_1\mathbf{X}_2}$ and $\overrightarrow{\mathbf{X}_1\mathbf{X}_3}$ is the altitude angle and can be computed as follows:
\begin{equation}
\label{eq:alt_angle_result1}
\begin{split}
	h &= \cos ^{-1} \frac{(\overrightarrow{\mathbf{X}_1\mathbf{X}_3})^T \overrightarrow{\mathbf{X}_1\mathbf{X}_2}} {\sqrt{(\overrightarrow{\mathbf{X}_1\mathbf{X}_3})^T \overrightarrow{\mathbf{X}_1\mathbf{X}_3}} \sqrt{(\overrightarrow{\mathbf{X}_1\mathbf{X}_2)}^T \overrightarrow{\mathbf{X}_1\mathbf{X}_2}}}\\
	&= \cos ^{-1} \sqrt{\frac{m_am_d}{m_am_d+m_bm_c}}\,.
\end{split}
\end{equation}
where the intermediate variables are as follows:
\[
	\begin{cases}
    	\begin{split}
    	m_a = & (f(x_1'-x_2')\sin\theta + (x_1'y_2'-x_2'y_1')\cos\theta)^2\\
    	& + f^2(y_2'-y_1')^2
    	\end{split}\\
    	m_b = {\cos}^4\theta (y_1'+f\tan\theta)^2 (y_2'+f\tan\theta)^2\\
        m_c = f^2(y_2'-y_3')^2\\
        m_d = (f\sin\theta+y_2'\cos\theta)^2 (f\cos\theta-y_3'\sin\theta)^2
	\end{cases}
\]
$x_i'=x_i-u_0$ and $y_i'=y_i-v_0$ for $i = 1, 2, 3$. The altitude angle $h$ should be in the range of $(0^{\circ}, 90^{\circ})$ when the sun is visible. 


\subsection{Estimating Sun Azimuth Angle}
\label{sec:azimuth_estimation}
To estimate the sun's azimuth angle $A$ from one shadow in the image, we design the following algorithm. The scenario is illustrated in Figure \ref{fig:one_shadow_scenario}. In particular, the point $\mathbf{X}_3$ is not necessary to be visible for estimating the azimuth angle. The true north $N$ is set to be the reference direction in our algorithm. The unit vector $\overrightarrow{I}=(0,0,1)$ has the same direction as the image direction.

The sun azimuth angle $A$ equals the angle measured clockwise around point $\mathbf{X}_1$ from due north to the shadow. We calculate $A$ as follows:
\begin{equation}
\label{eq:azimuth_angle}
	A=\angle (N, \overrightarrow{I}) + \angle (\overrightarrow{I},\overrightarrow{\mathbf{X}_1\mathbf{X}_2})\,,
\end{equation}
where $\angle (\overrightarrow{I},\overrightarrow{\mathbf{X}_1\mathbf{X}_2})$ denotes the angle measured clockwise from $\overrightarrow{I}$ to $\overrightarrow{\mathbf{X}_1\mathbf{X}_2}$, and $\angle (N, \overrightarrow{I})$ is the angle measured clockwise from $N$ to $\overrightarrow{I}$, which equals the orientation of the image direction. $\angle (\overrightarrow{I},\overrightarrow{\mathbf{X}_1\mathbf{X}_2})$ is the only unknown variable in Eq. \ref{eq:azimuth_angle}. 

We represent the angle between $\overrightarrow{I}$ and $\overrightarrow{\mathbf{X}_1\mathbf{X}_2}$ by $\alpha$. If $\angle (\overrightarrow{I},\overrightarrow{\mathbf{X}_1\mathbf{X}_2}) \leq 180^{\circ}$, it equals $\alpha$. Otherwise, it equals $(360^{\circ}-\alpha)$. Using the result of $\overrightarrow{\mathbf{X}_1\mathbf{X}_2}$ in last subsection,
we calculated the angle $\alpha$ as:
\begin{equation}
\label{eq:part_azimuth_angle}
\begin{split}
	\alpha & = \cos ^{-1} \frac{\overrightarrow{I}^T \overrightarrow{\mathbf{X}_1\mathbf{X}_2}} {\sqrt{\overrightarrow{I}^T \overrightarrow{I}} \sqrt{(\overrightarrow{\mathbf{X}_1\mathbf{X}_2)}^T \overrightarrow{\mathbf{X}_1\mathbf{X}_2}}}\\
	& = \cos ^{-1} \frac{m_a'} {\sqrt{{m_a'}^2 + {m_b'}^2}}\,,
\end{split}
\end{equation}
where
\[
	\begin{cases}
    	m_a' = f(y_2'-y_1')\\
    	m_b' = f(x_1'-x_2')\sin\theta + (x_1'y_2'-x_2'y_1')\cos\theta
	\end{cases}
\]

\section{Metadata-inferred Sun Position and Validation}
\label{sec:validation}

In this section, we describe the process to validate the consistency of a photo's capture time and location. The key idea is the following: we calculate the sun position using the capture time and location in the metadata of images. If the capture time and location are true, the sun position will match the one we estimated from shadows.

\subsection{Metadata-inferred Sun Position}
\label{subsub:calculate_sun_potion}
As mentioned in Section \ref{sec:background}, the position of the sun depends on the time of day, the date and the location of the observer. Its movement across the sky obeys the rules that have been studied in astronomy. In this section, we discuss the astronomical algorithms that are used to calculate \MSP{}, given the time and location.

We refer the time of day as the local time based on the standard time offsets of Coordinated Universal Time (UTC). However, the local standard time doesn't provide an intuitive connection with the sun position. In astronomy, the solar time is often used to discuss the sun position. It works because the sun finishes a $360^{\circ}$ rotation around the celestial sphere every 24 hours. The completed journey is divided into 24 hour, and one solar hour means that the sun travels a $15^{\circ}$ arc \cite{FundAstronomy}. The instant when the sun is due south in the sky or the shadow points to exactly north is called solar noon, which is 12:00 for solar time. Every $15^{\circ}$ arc the sun travels, one hour is added to 12:00 under the 24-hour clock system, and the angle distance that the sun passes on the celestial sphere is defined as the hour angle $H$ \cite{FundAstronomy}. It is measured from the sun's solar noon position, and ranges from $0^{\circ}$ to $+180^{\circ}$ westwards and from $0^{\circ}$ to $-180^{\circ}$ eastwards. The conversion between the local standard time $t_l$ to the solar time $t_s$ is as follows \cite{holbert2011solar, Sundials}:
\begin{equation}
t_s=t_l+ET+\frac{4 \ min}{deg}(\lambda_{std}-\lambda_{l})\,,
\end{equation}
where $\lambda_l$ denotes the local longitude, and $\lambda_{std}$ is the local longitude of standard time meridian, and $ET$ stands for the equation of time, which describes the difference of the true solar time and the mean solar time \cite{holbert2011solar}. The sun's hour angle is calculated as follows:
\begin{equation}
H=15^{\circ}(t_s-12)\,.
\end{equation}

Using the observer's local horizon as a reference plane, the azimuth and altitude angles of the sun can be calculated as follows \cite{Sundials}:
\begin{equation}
\label{eq:sun_azimuth_cal}
\tan (A) = \frac{\sin H}{\sin\varphi \cos H - \cos\varphi \tan\delta}\,,
\end{equation}

\begin{equation}
\label{eq:sun_altitude_cal}
\sin (h) = \sin\delta \sin\varphi + \cos\varphi \cos\delta \cos H\,,
\end{equation}
where $\varphi$ is the latitude of the observer's location, and $\delta$ is the sun's declination angle and it can be calculated as below \cite{iqbal2012introduction, Sundials}:
\begin{equation}
\delta = -23.44^{\circ}\cos(\frac{360^{\circ}(N+10)}{365^{\circ}})\,,
\end{equation}
where $N$ is the number of days since January 1st. Note that the azimuth angle $A$ calculated in Eq. \ref{eq:sun_azimuth_cal} uses south as a reference. We can derive the azimuth angle according to its definition in Section \ref{sec:background}.

\subsection{Consistency Validation}
\label{subsec:threshold}
Once obtaining the \SSP{} and \MSP{}, we check the difference between these two estimations by comparing their altitude angles and azimuth angles respectively. However, since there exists random and systemic errors in the \SSP{}, the estimation may not equal the ``true'' sun position. Thus, we have to select a threshold that is large enough to tolerate the errors yet small enough to detect the inconsistency between the \SSP{} and \MSP{}. Intuitively, the closer these two sun positions are to each other, the more likely the capture time and location are true.

We define the altitude angles of \SSP{} and \MSP{} to be $h_s$ and $h_m$ respectively, and the corresponding azimuth angles to be $A_s$ and $A_m$. Then the distance of the two altitude angles is $\mathrm{d_h}=|h_s-h_m|$, and the distance of the two azimuth angles is computed as $\mathrm{d_A}=|A_s-A_m|$. The likelihood of the consistency is inversely proportional to $\mathrm{d_h}$ and $\mathrm{d_A}$. However, the effects on $\mathrm{d_h}$ and $\mathrm{d_A}$ caused by fake capture time and/or location are different. For example, modifying the capture time from 12:00 p.m. to 13:00 p.m. may lead to $10^\circ$ in $\mathrm{d_A}$ but only $2^\circ$ in $\mathrm{d_h}$. So two different thresholds for $\mathrm{d_h}$ and $\mathrm{d_A}$ have to be selected. The capture time and location are considered to be true only when both $\mathrm{d_h}$ and $\mathrm{d_A}$ are within the thresholds. Besides, the sun position can be described by a pair of azimuth angle and altitude angle: $(A, h)$. We can also use the sun position distance that is computed as $\mathrm{d_p}=\sqrt{\mathrm{d_A}^2+\mathrm{d_h}^2}$ to distinguish the two estimations of the sun position. Our goal is to choose appropriate variables and thresholds that can increase the probability of correct validation for inconsistent images and decrease the probability of false validation for consistent images. Section \ref{sec:results} details the selection of thresholds in the validation experiment.

\section{Evaluation}
\label{sec:results}

This section presents the results of our experiments. To evaluate the performance of the \SSP{} algorithms, we built a computer synthetic dataset and conducted extensive tests with it. To validate the effectiveness of the framework \framework{}, we gathered 200 photos in China, the U.S. and Japan in the span of about one year, and examined whether we can detect the modifications of capture time, date and location.

\subsection{Datasets}
\label{sec:datasets}
We build two datasets for evaluating our algorithms and framework.
 
\subsubsection{Dataset \Rmnum{1}: synthetic photos}
\label{sec:dataset1}
We choose the city of New York, the 21st day of March 2017, and the time period from 9:00 a.m. to 4:00 p.m. to simulate the sun positions. We generate a photo of a synthetic object and its shadow every 5 minutes using the following simulated camera model. 

\textbf{Camera model.} As shown in Figure \ref{fig:one_shadow_scenario}, we build a pinhole camera model using the open source package: OpenCV~\footnote{http://opencv-python-tutroals.readthedocs.io/en/latest/index.html}. The simulated camera has the following parameters: the focal length $f = 3351.6$ pixels (equivalent to $3.99$mm based on a $4.8$mm wide CCD), the principle point $(u_0, v_0) = (2016, 1512)$ pixels. As expected with most modern CCD cameras, it has zero skewness and unit aspect ratio. We also set the camera height above the ground plane to be 1.6 meters ($1$m $\equiv 840000$ pixels) which is close to the real height at which people hold a camera to take photos.

\textbf{Scene setting.} We place one vertical object on the ground plane along the axis of $\mathrm{Z}_w$ (refer to \fig \ref{fig:one_shadow_scenario}). The object has a height of $1$ meter and a distance of $m$ meters from the camera. Mathematically, the object's footprint $\mathbf{X}_2$ is at ($0, 1.6, m$) meters for $m=5, 10$ in the world coordinate system. Under each scenario, we generate 84 photos of the synthetic object and its shadow.

\begin{figure*}[t]
\captionsetup[subfigure]{justification=centering}
\centering

\null\hfill
\begin{subfigure}{0.45\textwidth}
	\includegraphics[width=\textwidth]{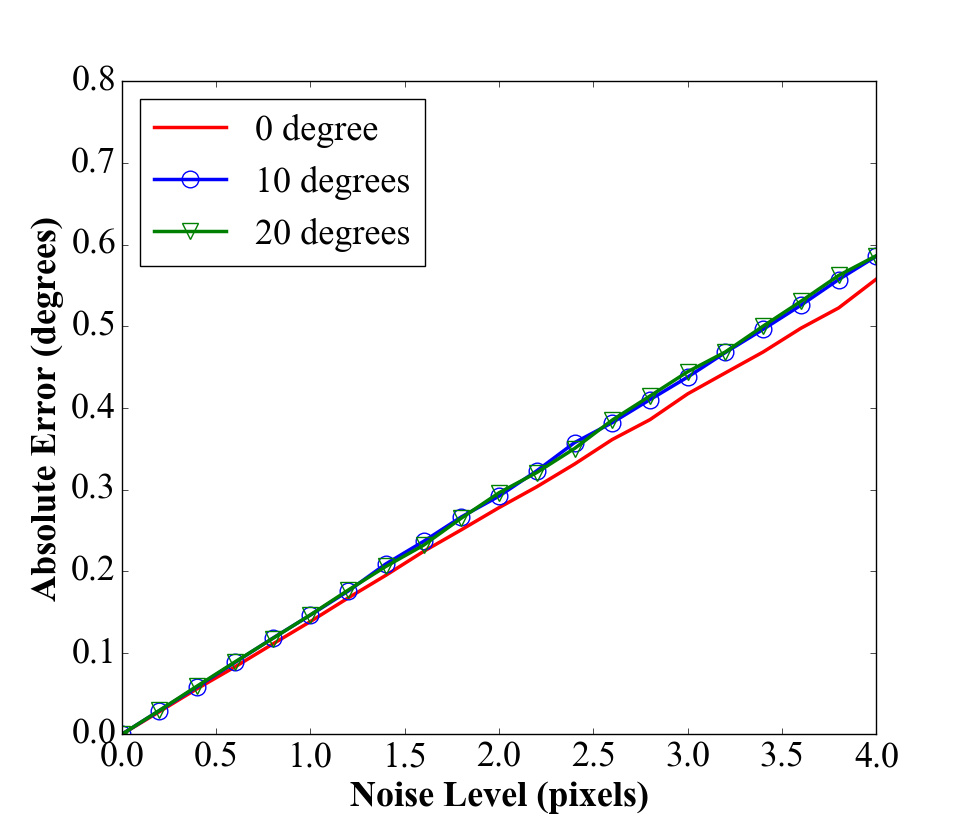}
	\caption{Absolute errors in sun altitude angles.}
	\label{fig:noise1_alt_alg1}
\end{subfigure}
\hfill
\begin{subfigure}{0.45\textwidth}
	
	\includegraphics[width=\textwidth]{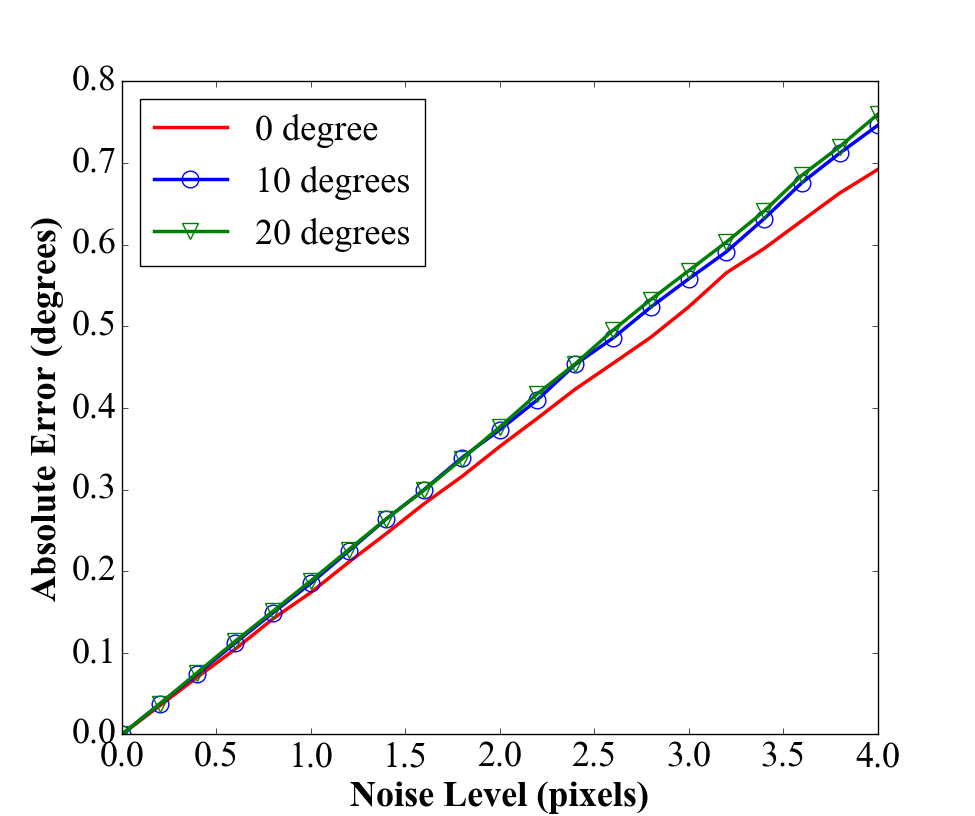}
	\caption{Absolute errors in sun azimuth angles.}
	\label{fig:noise1_azi_alg1}
\end{subfigure}
\hfill\null
\\
\null\hfill
\begin{subfigure}{0.45\textwidth}
	
	\includegraphics[width=\textwidth]{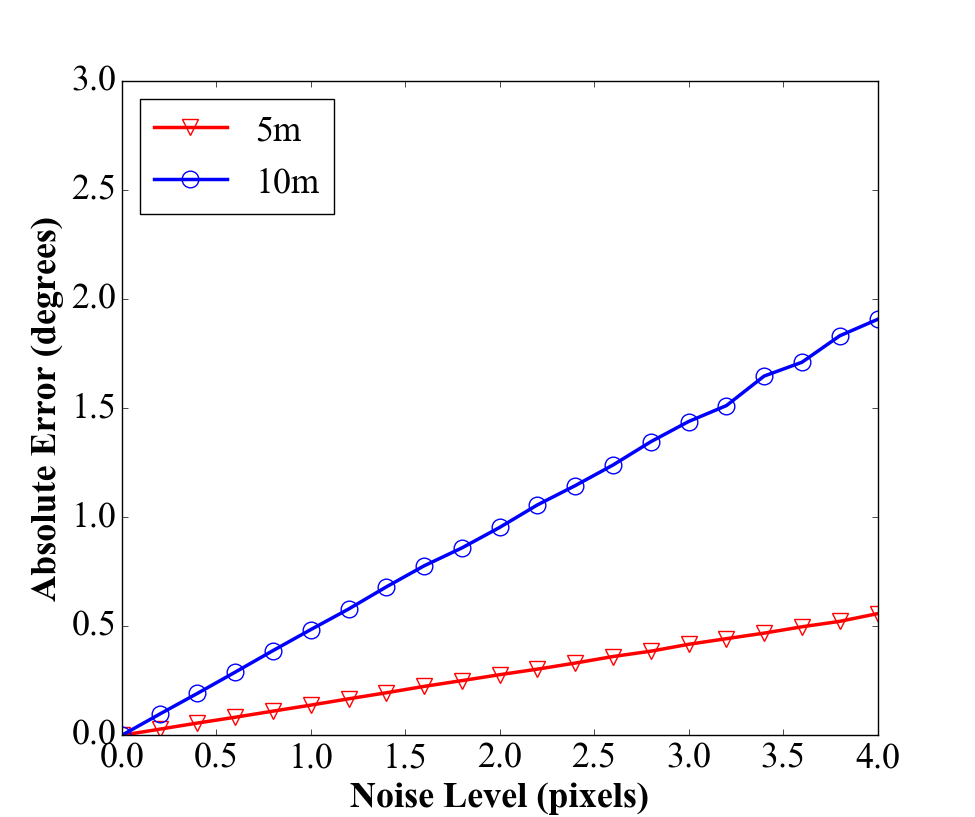}
	\caption{Absolute errors in sun altitude angles.}
	\label{fig:noise2_alt_alg1}
\end{subfigure}
\hfill
\begin{subfigure}{0.45\textwidth}
	
	\includegraphics[width=\textwidth]{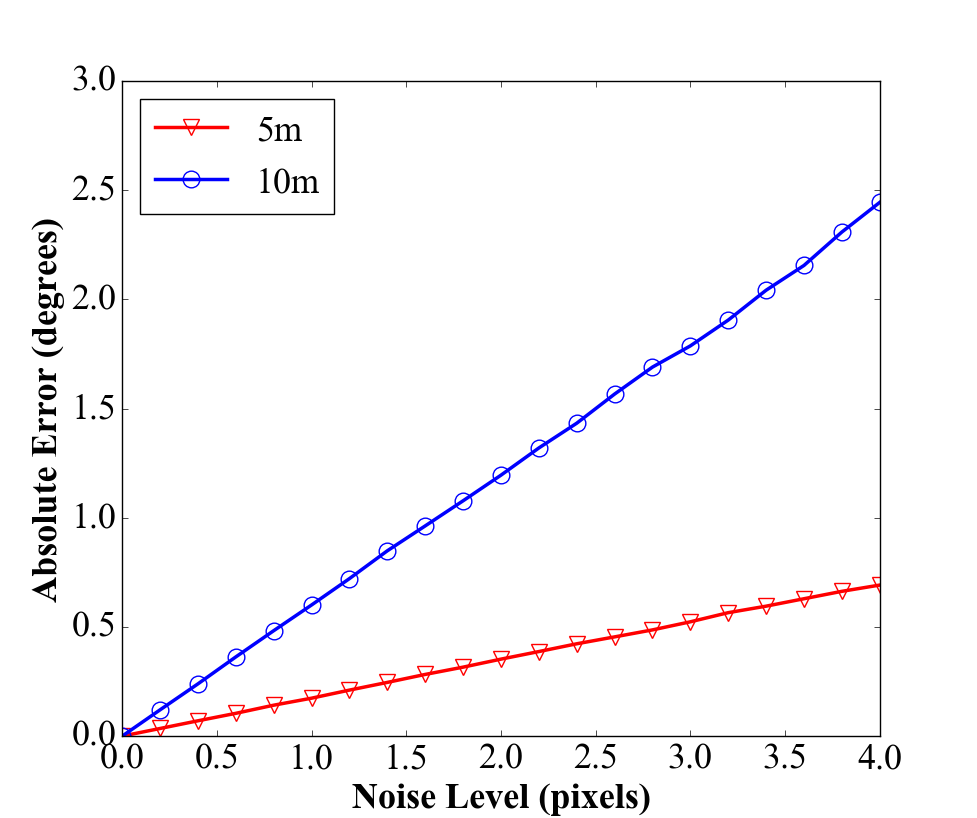}
	\caption{Absolute errors in sun azimuth angles.}
	\label{fig:noise2_azi_alg1}
\end{subfigure}
\hfill\null
\caption{Performances on synthetic data containing Gaussian noise: Figure \ref{fig:noise1_alt_alg1} and \ref{fig:noise1_azi_alg1} depict the absolute errors in sun positions varying with the camera inclination angles, while Figure \ref{fig:noise2_alt_alg1} and \ref{fig:noise2_azi_alg1} show the absolute errors in sun positions varying with the distances of the object and camera.}
\label{fig:syn_data_noise}
\end{figure*}

\subsubsection{Dataset \Rmnum{2}: real photos}
\label{sec:dataset2}
This dataset consists of 200 photos taken by 10 different smartphones, including iPhone 5s, 6, 6 plus, 6s, 6s plus and 7. They were captured at 16 cities around China, the U.S. and Japan since September 2016.  Among the 200 photos, 79 were taken in China, 7 were captured in Japan, and the rest were collected in the U.S. Each photo encloses the metadata that includes the real capture time and location. In addition, 76 out of the 200 photos were taken by cameras tilted forwards or backwards with different angles. 
Our dataset mainly contains three types of vertical objects: standing people, poles (e.g. road signs, lampposts) and tree trunks. These objects are frequently seen in the outdoor photos and are almost vertical to the ground. 

We refer to the true metadata of the 200 photos as the \textit{positive samples}. We generate the attack data by falsifying the metadata of the 200 photos and refer to the attack metadata as the \textit{negative samples}. Note that multiple types of metadata may result in the same effect. For instance, modifying longitude one degree more to the west has the same effect on the sun position as changing the local time forward by four minutes. Thus, falsifying either longitude or the local time is equivalent. To simplify the analysis yet without loss of generality, we focus on three types of attacks that modify the following metadata:
\begin{itemize}
  \item Falsified time of day, and true date and location.
  \item Falsified date, and true time and location.
  \item Falsified latitude of location, and true time and date.
\end{itemize}

The ``fake'' times of day are randomly generated in the range from 8:00 a.m. to 17:00 p.m. when the sun is likely to be seen. The ``fake'' dates are randomly generated from the range within one year. The ``fake'' latitudes of location are randomly generated in the range of $25^\circ$ and $50^\circ$ of the Northern Hemisphere where most parts of the U.S., Japan and China locate. We have 200 negative samples for each type of attack metadata.

\subsection{Evaluation of the Shadow-Inferred Sun Position Algorithms}
\label{sec:algo_evaluation}

Using Dataset \Rmnum{1}, we mainly examine the accuracy of the \SSP{} algorithms and their resilience to noise. We also analyze the impact of varying the camera inclination angle and the distance between the camera and object on estimating sun positions. 

When applying our algorithms to the scene setting described in Section \ref{sec:dataset1}, we add Gaussian noise to the interested object and shadow points (\ie $\mathbf{X}_1, \mathbf{X}_2, \mathbf{X}_3$). We refer to variance/covariance as the noise level. For each noise level, we performed 200 independent trails on each photo (84 photos in total). Then, we compute the absolute error with respect to the ground truth for each trail. Finally, we calculate the average absolute error of the 84 photos for each noise level.

To plot \fig \ref{fig:noise1_alt_alg1} and \ref{fig:noise1_azi_alg1}, we set the object to be $5$ meters away from the camera and vary the camera inclination angle ($\theta = 0^{\circ}, 10^{\circ}, 20^{\circ}$); To plot \fig \ref{fig:noise2_alt_alg1} and \ref{fig:noise2_azi_alg1}, we set the camera inclination angle to be $0^{\circ}$ and vary the distance ($5, 10$ meters) between the camera and the object.

\textbf{Noise resilience.} \fig \ref{fig:syn_data_noise} indicates that the absolute error increases almost linearly as the noise level grows. We obtain the maximum absolute error of $1.91^{\circ}$ in sun altitude angles (\fig \ref{fig:noise2_alt_alg1}) and $2.44^{\circ}$ in sun azimuth angles (\fig \ref{fig:noise2_azi_alg1}) when the noise level reaches 4 pixels and the object is $10$ meters away from the camera. The absolute errors in sun altitude angles (\fig \ref{fig:noise1_alt_alg1}, \ref{fig:noise2_alt_alg1}) are slightly smaller than the errors in sun azimuth angles (\fig \ref{fig:noise1_azi_alg1}, \ref{fig:noise2_azi_alg1}) under the same setting of parameters. In short, \fig \ref{fig:syn_data_noise} demonstrates that both algorithms can achieve good accuracy even in the high noise case (4 pixels), and the algorithm for estimating sun altitude angle provides a slightly stronger resilience to noise than the algorithm for estimating sun azimuth angle.

\textbf{Impact of different camera inclination angles.} \fig \ref{fig:noise1_alt_alg1} and \ref{fig:noise1_azi_alg1} show that varying the camera inclination angle has little effect on the noise resilience of both algorithms. As the inclination angle increases, the absolute error in either the sun altitude angle or the azimuth angle only increases a little bit at the same noise level. Our algorithms are robust to the camera inclinations

\textbf{Impact of different distances from the camera to objects.} \fig \ref{fig:noise2_alt_alg1} and \ref{fig:noise2_azi_alg1} denote that increasing the distance between the camera and object has a negative impact on the noise resilience of both algorithms. As the distance increases, the absolute error grows. The error can be large if the noise level is very high and the distance is long. Thus, our algorithms can achieve better performance when the distance of the camera and the interested object is short (\ie $<10$ m). Compared to the camera inclination angle (\fig \ref{fig:noise1_alt_alg1}, \ref{fig:noise1_azi_alg1}), our algorithms are more sensitive to the distance (\fig \ref{fig:noise2_alt_alg1}, \ref{fig:noise2_azi_alg1}).

In summary, our algorithms in Section \ref{sec:estimate_sun_position} are able to infer sun position with high accuracy and robust to noise.

\subsection{Evaluation of $\protect\framework$}
To evaluate the performance of \framework{} and to understand threshold selection, we conducted a set of experiments.

\subsubsection{Metric}
We use ROC curves to evaluate the performance of \framework{} by varying thresholds for our system. An ROC curve represents Receiver Operating Characteristic curve and is created by plotting true positive rate (TPR) against false positive rate (FPR), as the threshold varies \cite{fawcett2006introduction}. The true positive rate and false positive rate are defined as below.
\begin{align*}
	\text{TPR} &= \frac{\# \ \textrm{of} \ \textrm{true} \ \textrm{positives}} {\# \ \textrm{of} \ (\textrm{true} \ \textrm{positives} + \textrm{false} \ \textrm{negatives})} \\
	\text{FPR} &= \frac{\# \ \textrm{of} \ \textrm{false} \ \textrm{positives}}{\# \ \textrm{of} \ (\textrm{true} \ \textrm{negatives} + \textrm{false} \ \textrm{positives})}	
\end{align*}
where a true positive denotes the result that a positive sample is correctly identified as such, and a false positive is the one that a negative sample is identified as a positive sample by mistakes. The point $(0,1)$ on the ROC curve denotes 0 FPR and 100\% TPR, which indicates an ideal system that can correctly identify all genuine photos and reject all falsified photos \cite{tian2013kinwrite}. In our experiment, we select the optimal threshold as the one that yields the minimum distance from the corresponding point on the ROC curve to the ideal point $(0,1)$. Another indicator that we use to evaluate the average performance of the validation is the area under the ROC curve (AUC). The closer it is to 1, the better the average performance is \cite{fawcett2006introduction}.


\begin{figure*}[t]
\captionsetup[subfigure]{justification=centering}
\centering

\null\hfill
\begin{subfigure}{0.32\textwidth} 
	\includegraphics[width=\textwidth]{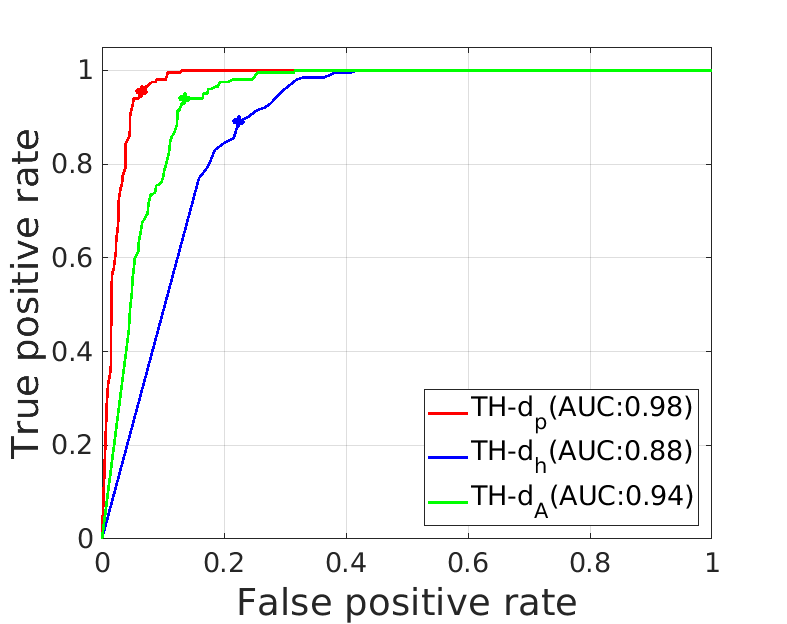}
	\caption{The attack metadata with falsified time of day.}
	\label{fig:roc_curveA}
\end{subfigure}
\hfill
\begin{subfigure}{0.32\textwidth}
	\includegraphics[width=\textwidth]{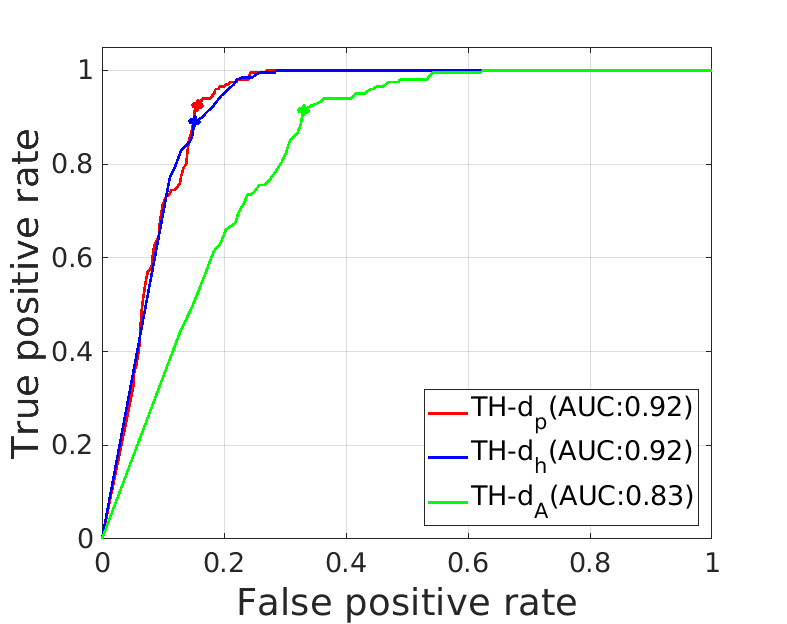}
	\caption{The attack metadata with falsified date.}
	\label{fig:roc_curveB}
\end{subfigure}
\hfill
\begin{subfigure}{0.32\textwidth}
	\includegraphics[width=\textwidth]{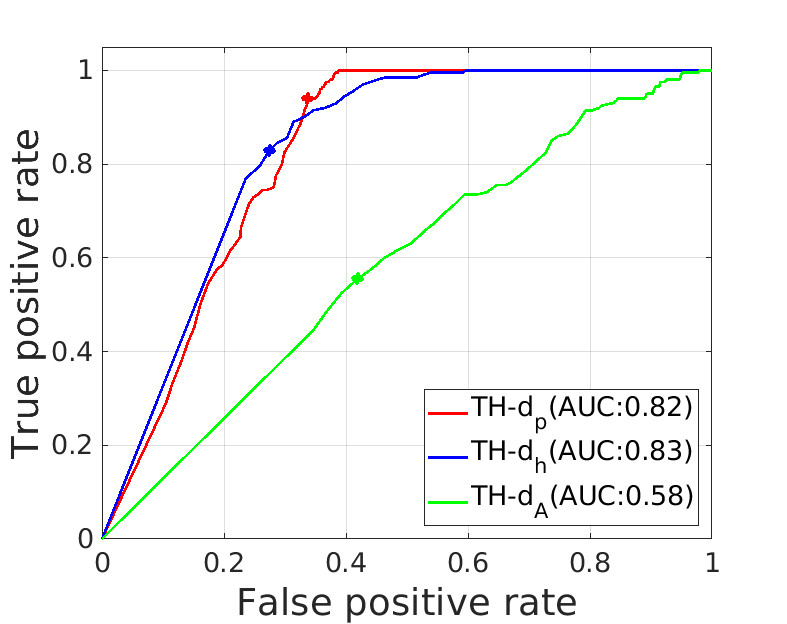}
	\caption{The attack metadata with falsified latitude.}
	\label{fig:roc_curveC}
\end{subfigure}
\hfill\null
\caption{ROC curves based on different distance variables and different types of attack metadata.}
\label{fig:roc_curve}
\end{figure*}

\begin{figure}
	\includegraphics[width=0.44\textwidth]{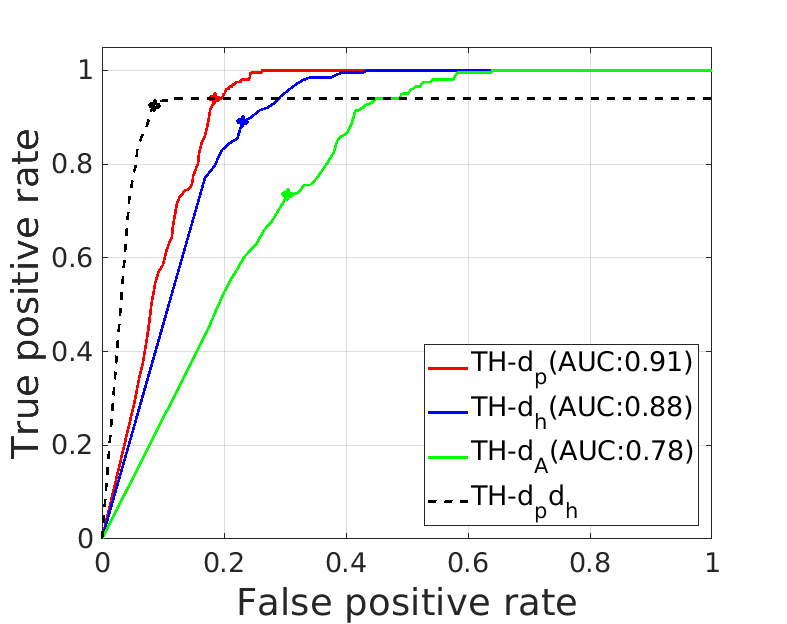}
	\caption{ROC curves for a collective of the three types of attack metadata.}
	\label{fig:roc_curve_all}
\end{figure}

\subsubsection{Performance and Thresholds} 
Based on the framework \framework{}, we performed consistency validation using the three types of falsified metadata in Dataset \Rmnum{2}. To understand how the altitude angle and azimuth angle influence the performance of the validation, we examine three distances separately: the distance of the altitude angles $\mathrm{d_h}$, the distance of the azimuth angles $\mathrm{d_A}$, and the distance of the sun positions $\mathrm{d_p}$. Here, the sun position is defined to be $(A,\, h)$, in which $A$ refers to the azimuth angle and $h$ refers to the altitude angle. To decide the best distance variable which can yield the maximum AUC and the optimal threshold of the variable, we analyze the ROC curves that are plotted by varying the threshold of each type of distance.

The results are presented in the set of ROC curves shown in Figure \ref{fig:roc_curve} and Figure \ref{fig:roc_curve_all}. Each ROC curve with distinct color is plotted by varying the threshold of one type of the three distances. ``$\mathrm{TH}$-$\mathrm{d_h}$'' and ``$\mathrm{TH}$-$\mathrm{d_A}$'' denote varying the threshold of the altitude angle distance $\mathrm{d_h}$ and the azimuth angle distance $\mathrm{d_A}$ respectively. ``$\mathrm{TH}$-$\mathrm{d_p}$'' denotes varying the threshold of the sun position distance $\mathrm{d_p}$. For each type of attack metadata, we repeat the random generating of the 200 negative samples 5 times. And each false positive rate on the ROC curve is averaged over these repeated attack metadata. 

Figure \ref{fig:roc_curveA} indicates that the detection based on $\mathrm{d_A}$ slightly outperforms the one based on $\mathrm{d_h}$. However, the $\mathrm{d_h}$ based detection achieves better performance in detecting all the other types of attacks as shown in Figure \ref{fig:roc_curveB} and Figure \ref{fig:roc_curveC}, especially in detecting falsified latitude. The result implies that $\mathrm{d_h}$ is more important in distinguishing different positions of the sun compared to $\mathrm{d_A}$ in general. Such a conclusion confirms with the result reported in Section \ref{sec:algo_evaluation}, i.e., the average estimation error of the altitude angles is smaller than that of the azimuth angles. If only $\mathrm{d_h}$ is used for consistency validation, Figure \ref{fig:roc_curve_all} guides us to choose the optimal threshold of $\mathrm{d_h}$ to be $3.2^\circ$ and it achieves combined (TPR, FPR) values of $(89\%, \ 23\%)$ for all attacks, which means that 89\% of positive samples can be correctly validated but 23\% of negative samples will be mistakenly identified. 

In addition, Figure~\ref{fig:roc_curve} shows that the $\mathrm{d_p}$ based detection achieves the best performance in detecting falsified time of day, and has almost the same performance as the $\mathrm{d_h}$ based detection in detecting the other types of attacks. Once we only use $\mathrm{d_p}$ for consistency validation, Figure \ref{fig:roc_curve_all} leads us to choose the optimal threshold of $\mathrm{d_p}$ to be $9.4^\circ$, which achieves combined (TPR, FPR) values of $(94\%, \ 18.4\%)$ for all attacks.

To improve the performance further, we examine both the $\mathrm{d_p}$ and $\mathrm{d_h}$ to validate the consistency of time and location. That is, a sample has to satisfy both the thresholds of $\mathrm{d_h}$ and $\mathrm{d_p}$ to be accepted by \framework{}. Plotting the ROC curves and finding the global optimal thresholds by varying two thresholds can be tricky. Thus, we chose the local optimal threshold for one variable and varied the other threshold to plot the ROC curve. This approach may not generate the global optimal thresholds for the two variables, but it strikes a balance between the optimum and the computational cost. We chose the threshold of $\mathrm{d_p}$ to be $9.4^\circ$ and varied the threshold of $\mathrm{d_h}$. The resulting curve illustrates an improved performance than the one of using a single threshold as shown in Figure \ref{fig:roc_curve_all}. Note that we cannot plot an integral ROC curve when the threshold of $\mathrm{d_p}$ is fixed since the highest true positive rate will be decided by the fixed threshold, which is 94\%. The curve ``$\mathrm{TH}$-$\mathrm{d_pd_h}$'' in Fig \ref{fig:roc_curve_all} indicates that choosing the optimal threshold of $\mathrm{d_h}$ to be $5^\circ$ can correctly identify 92.5\% positive samples, but cannot identify 8.5\% of negative samples.

\section{Discussions}
\label{sec:discussions}
In this section, we discuss the attacks against our framework and analyze two factors that may influence the estimation of sun position.

\subsection{Attacks against \framework{}}
\label{sec:attacks_on_framework}
Based on the evaluation results, we analyze the robustness of the framework \framework{} when detecting the falsifications of the time of day, the date and the location. \framework{} can confidently identify the attacks that cause violations of one of the thresholds of the altitude angle distance and the sun position distance. We provide the following examples to help have an intuitive understanding of how much the time and location change may result in an exceeding of the chosen thresholds. 

First, we assume the location is New York city and it is not modified. We choose three representative dates (e.g., the 21st of December, March and June) when the sun looks lowest, intermediate and highest, respectively, to check how much modification of the time can be detected by \framework{}. We choose the time of 9:00 a.m. and 12:00 p.m. as the baselines. Table \ref{tab:time_drift} shows the results. The sun appears to move fastest on June 21st and a modification more than 16 minutes from 12:00 p.m. will exceed the sun position threshold, while the sun moves slowest on December 21st and a modification greater than 40 minutes can be recognized. 

Second, we assume both the time of the day and the location are true. We vary the date of the year for validations. We use the 21st of December, March and June as our baselines. As shown in Table \ref{tab:data_drift}, \framework{} can correctly detect a change greater than 13 days from March 21st at 12:00 p.m. and at worst a 48-day change from June 21st at 9:00 a.m. The reason that the difference exists is because the sun's daily path does not change at a constant rate throughout the year.

Third, we fix the time and the date but modify the location. We find that \framework{} is able to detect location modifications that are larger than $400$ miles away from New York. There is another scenario: an attacker knows how to modify both the time and location of a photo such that the altitude angle and the sun position are within the thresholds and the modification can fool \framework{}. Luckily. the motivation of falsifying the metadata of a photo is to use it for a chosen event and the attacker may not be guaranteed to find such a combination.


\begin{table}[t]
  \centering
    \caption{Time modifications that can be detected.}
  \begin{tabular}{ccc}
    \hline
	\multirow{2}{*}{\textbf{Date}}  &  \multicolumn{2}{c}{$\Delta$\textbf{T} (minutes)} \\ 
	  &  9:00 a.m.  &  12:00 p.m. \\ \hline 
    Dec. 21st  &  $\geq40$  &  $\geq38$ \\
    Mar. 21st  &  $\geq29$  &  $\geq26$ \\
    June 21st  &  $\geq27$  &  $\geq16$ \\ \hline   
  \end{tabular}
  \label{tab:time_drift}
\end{table}

\begin{table}[t]
  \centering 
    \caption{Date modifications that can be detected.}
  \begin{tabular}{cccc}
    \hline
	\multirow{2}{*}{\textbf{Time}}  &  \multicolumn{3}{c}{$\Delta$\textbf{D} (days)} \\ 
	  &  Dec. 21st  &  Mar. 21st  &  June 21st \\ \hline
    9:00 a.m.  &  $\geq33$  &  $\geq16$  &  $\geq48$ \\
    12:00 p.m.  &  $\geq32$  &  $\geq13$  &  $\geq40$ \\ \hline   
  \end{tabular}
  \label{tab:data_drift}
\end{table}

Referring to the fake photo given in Section \ref{sec:introduction}, we show how \framework{} detects that it was not taken at the claimed dates and locations. For the image in Figure \ref{fig:fake_explosion}, although we do not have the required metadata information (e.g., the camera orientation) to estimate the azimuth angle as well as the exact sun position, we can estimate the altitude angle range. Given the claimed time (i.e., 16:50 on June 15, 2017) and the location (i.e., Xuzhou, China), we can calculate the sun altitude angle to be $29.1^\circ$ based on astronomical algorithms. On the other hand, based on the image, we estimate the focal length to be $420 \pm 20$ pixels using algorithms in \cite{geotempestimation}. By assuming the camera inclination angle to be between $-10^\circ$ and $10^\circ$, we obtain the altitude angle to be between $52^\circ$ and $58^\circ$. The distance between the two estimates is far beyond the threshold $5^\circ$ in our experiments. Thus, we conclude that the date and location of this image were spoofed.

\subsection{Influence of Ground Slope}
\label{sec:ground_slope}
One source of uncertainty concerns the possibility of ground slope which is associated with the accuracy in locating the ``real'' shadow's position. If the ground where the shadow located is not truely flat and has a slope angle of $\Delta G$ with respect to the horizontal plane, $\Delta G$ will propagate as the altitude angle is estimated. It also causes a slight error in the estimated azimuth angle. But neither one of the errors will be larger than $\Delta G$. On the other hand, if our goal is to verify the capture time of a photo (i.e., the capture location has been identified through other methods), we can measure the slope angle by visiting the place and use it to calibrate the estimated sun position. Based on the preciser estimation, our framework will be more confident in validating the capture time of the image.

\subsection{Influence of Non-vertical Object}
\label{sec:nonvertical_obj}
Possible tilt angle of the ``vertical'' object is another source of uncertainty. Although many well-structured man-made objects (e.g., road signs, lampposts, walls) usually include strong vertical edges with respect to the ground\cite{VAILAYA19981921, zhang15geoscience, Lee14pami}, some certain errors may exist. Upright standing human beings in images may also not be perfectly vertical to the ground. The tilt angle will mainly cause error in the estimated sun altitude angle and have little effect on the shadow orientation. The resulting error will not be larger than the tilt angle itself. The maximum tilt angle is $90^\circ$ when the object tilt on the ground, and it causes an error equal to the true sun altitude angle which is less than $90^\circ$. On the other hand, if we can locate where the image was taken, the estimated altitude angle can be calibrated by measuring the tilt angle of the object. Then we are able to improve the confidence of our framework in validating the capture time. 

\section{Related Work}
\label{sec:related_work}
In the field of computer vision, there have been many studies aimed at estimating the geolocation or the time of cameras and images. Studies have demonstrated that shadow trajectories inferred from multiple outdoor images can be used to determine the geolocation of stationary cameras~\cite{geotempestimation, Wu2010}. Sandnes~\cite{Sandnes2011} uses the relative lengths of objects and their shadows in images to estimate the sun altitude angle and further estimates the geolocation of the camera. Jacobs \etal~\cite{jacobs07geolocate} analyze the correlations of camera images with satellite images and other images with known locations to determine where these camera images were taken. 

Researchers have also used sun position to estimate the capture time of images. Based on Lalonde \etals model~\cite{Lalonde2009, Lalonde2010} that calculates  the sun position by combining the cues in the image such as the sky and the shadows on the ground, Tsai \etal~\cite{Tsai2016529} and Kakar \etal~\cite{KakarS12a} utilize the position of the sun to estimate the capture time for geo-tagged images. Other studies extract fluctuations of the Electric Network Frequency (ENF) signal from fluorescent lighting and verify the feasibility of using the ENF signal as a natural timestamp for videos in an indoor enviroment~\cite{garg2011seeing, Chuang_2012}. Deep learning algorithms have recently been used to predict the contextual information of outdoor images. Volokitin \etal~\cite{Volokitin_2016_CVPRW} apply convolutional neural networks for the predictions of temperature and time of the year. 

Chen \etals~\cite{chen_cvprw} work is probably most related to ours. The authors train machine learning models to predict the sun altitude angle, temperature, humility and weather condition associated with outdoor images, and then combine all the inferred information to detect tampering in the image metadata. Their results show that the absolute error in predicting sun altitude angle is larger than $10^{\circ}$ for $45\%$ of the testing photos. Our method can compute both sun altitude angle and azimuth angle without requiring a large amount of photos for training models. Moreover, our algorithms can achieve lower error ($2.98^{\circ}$) in estimating sun altitude angle.

\section{CONCLUSION}
\label{sec:conclusion}
We presented a new framework \framework{} which uses two estimations of sun position\textemdash{}\SSP{} and \MSP{}\textemdash{}to check whether the capture time and location of an outdoor image are true. Our framework exploits the relationship between the sun position in the sky and the time and location of an observer. We designed algorithms to obtain \SSP{} using vertical objects and their shadows in the image. Our experiments show that the algorithms can estimate the sun position from shadows in the image with satisfactory accuracy.  The evaluation results demonstrate that \framework{} can achieve combined (TPR, FPR) values of $(92.5\%, \ 8.5\%)$ for the consistency validation. We believe that our results illustrate the potential of using sun position to validate the consistency of the capture time and location. Our work raises an open question that whether other image contents can be leveraged for validating the consistency of image's contextual information.



\appendices

\section{Proof of Equation \ref{eq:alt_angle_result1}}
\label{app:proof_of_first_eq}
To calculate the coordinates of $\mathbf{X}_1$ and $\mathbf{X}_2$, we have the following equations:
\begin{subequations}
\label{eq:pointsX1X2}
\begin{align}
        \mathbf{X}_i^T \mathbf{G} &= 0,\\
        \mathbf{x}_i &= \mathrm{P}\mathbf{X}_i,
\end{align}
\end{subequations}
where $i = 1,2$ and $\mathbf{G}=[0, 1, 0, h_c]^T$. By solving the above equations, we can obtain the coordinates of points $\mathbf{X}_1$ and $\mathbf{X}_2$. Then we have the vector $\overrightarrow{\mathbf{X}_1\mathbf{X}_2}$. We define $\mathbf{x}_i$ to be $[x_i,y_i,1]$ under the image coordinate frame, and $\mathbf{X}_i$ to be $[X_i,Y_i,Z_i,1]$ under the world coordinate frame.

\begin{equation}
	\mathbf{X}_i = 
\begin{bmatrix}
h_c \frac{-x_i'} {\cos\theta (y_i' + f\tan\theta)}\\
-h_c\\
h_c \frac{y_i'\tan\theta - f} {y_i' + f\tan\theta}\\
1
\end{bmatrix}\quad i = 1,2 ,
\end{equation}
where $x_i'=x_i-u_0$ and $y_i'=y_i-v_0$. Then we have the vector 
\begin{equation}
\label{eq:vector_x1x2}
\overrightarrow{\mathbf{X}_1\mathbf{X}_2} = 
\begin{bmatrix}
\frac{f(x_1'-x_2')\tan\theta+(x_1'y_2'-x_2'y_1')} {\cos\theta(y_1'+f\tan\theta)(y_2'+f\tan\theta)}\\
0\\
\frac{f(y_2'-y_1')}{{\cos}^2\theta(y_1'+f\tan\theta)(y_2'+f\tan\theta)}
\end{bmatrix} .
\end{equation}
Note that we divide each resulting coordinate by $h_c$ in Eq. \ref{eq:vector_x1x2}.

\begin{equation}
	\mathbf{X}_3 = 
\begin{bmatrix}
{X}_2\\
\frac{{Z_2}(f\sin\theta + y_3'\cos\theta)} {f\cos\theta - y_3'\sin\theta}\\
{Z}_2\\
1
\end{bmatrix} .
\end{equation}
Using the coordinates of $\mathbf{X}_1$ and $\mathbf{X}_3$, we can calculate the vector
\begin{equation}
\overrightarrow{\mathbf{X}_1\mathbf{X}_3} = 
\begin{bmatrix}
\frac{f(x_1'-x_2')\tan\theta+(x_1'y_2'-x_2'y_1')} {\cos\theta(y_1'+f\tan\theta)(y_2'+f\tan\theta)}\\
\frac{f(y_2'-y_3')} {(f\sin\theta+y_2'\cos\theta)(f\cos\theta-y_3'\sin\theta)}\\
\frac{f(y_2'-y_1')} {{\cos}^2\theta(y_1'+f\tan\theta)(y_2'+f\tan\theta)}
\end{bmatrix}.
\end{equation}

%
%
%
%
%
%

\bibliographystyle{IEEEtran}
\bibliography{references}

%







\end{document}